\documentclass{article}
\usepackage{amsmath}
\usepackage{amstext}
\usepackage{amssymb}
\usepackage{graphicx}
\usepackage{fullpage}
\title{White Dwarf based evaluation of the GALEX absolute calibration}
\author{L. Camarota,$^{1,2}$ \&J. B. Holberg$^{1,3}$
\\
\footnotesize $^{1}$Lunar and Planetary Laboratory, University of Arizona, Sonett Space Sciences Building
\\
\footnotesize 1541 East University Boulevard, Tucson, AZ 85721-0063
\\
\footnotesize $^{2}$camarota@physics.arizona.edu
\\
\footnotesize$^{3}$holberg@argus.lpl.arizona.edu}
\begin{document}
\label{firstpage}
\maketitle
\begin{abstract}
\normalsize This paper describes a revised photometric calibration of the \emph{Galaxy Evolution Explorer} magnitudes, based on measurements of DA white dwarfs.  The photometric magnitudes of white dwarfs measured by \emph{GALEX} are compared to predicted magnitudes based on independent spectroscopic data (108 stars) and alternately to \emph{IUE} UV fluxes of the white dwarfs (218 stars).  The results demonstrate a significant non-linear correlation and small offset between archived \emph{GALEX} fluxes and observed and predicted UV fluxes for our sample.  The primary source of non-linearity may be due to detector dead time corrections for brighter stars, but it should be noted that there was a predicted non-linearity in the fainter stars as well.  Sample expressions are derived which 'correct' observed \emph{GALEX} magnitudes to an absolute magnitude scale that is linear with respect, and directly related, to the \emph{Hubble Space Telescope} photometric scale.  These corrections should be valid for stars dimmer than magnitudes 9.3 and 10.5 in the NUV and FUV respectively, and brighter than magnitude 17.5 in both
\end{abstract}
\pagebreak

\section{Introduction}

\subsection {\emph{GALEX} Mission}
The \emph{Galaxy Evolution Explorer} (\emph{GALEX}) was a NASA Small Explorer class mission launched in 2003, whose primary objective was to conduct deep ultraviolet surveys of the sky in two broadband bands between 1400\AA\ and 3000\AA\ in order to study the stellar evolution of faint external galaxies (Bianchi et al. 2000).  It was operated by the California Institute of Technology with NASA support until 2012.  Thereafter it was supported by Caltech with a private fund raising effort called \emph{Complete the All-sky UV Survey Extension}, or \emph{CAUSE}.  The \emph{GALEX} mission was decommissioned in 2013.
\\
\\
\emph{GALEX} operated primarily in a pointed mode, which tiled the sky with circular fields approximately 1.2$^{\circ}$ in diameter.  Approximately 75 per cent of the sky was observed during the \emph{GALEX} mission.  Areas near UV-bright sources and near the Galactic plane and other crowded regions were generally avoided by these observations.  When the California Institute of Technology assumed operational control, some observation parameters were changed.  \emph{GALEX} started using a scanning mode to record areas that were brighter than previously permitted, including approximately 80$^{\circ}$ of the Galactic plane.
\\
\\
The primary mission of \emph{GALEX} was to conduct observations pertaining to the spectral evolution of galaxies in the UV. The majority of \emph{GALEX} observations involved four surveys.  The Deep Imaging Survey (DIS) made long exposure (30,000 s) observations of regions with targets that had been extensively observed in other wavelength bands.  The Medium Imaging Survey (MIS) made single orbit exposure (1,500 s) observations of regions with targets observed by the Sloan Digital Sky Survey (SDSS) spectroscopic footprint, the Two Degree Field Galaxy Redshift Survey (2dFGRS), and the AA-Omega (WiggleZ) project.  The All sky Imaging Survey (AIS) made short exposure (100s) observations of as much of the sky as possible.  The Nearby Galaxy Survey (NGS) made single orbit exposure (1,500 s) observations of nearby galaxies that have significant observations in other wavelengths.  \emph{ GALEX} also devoted 33 per cent of its observation time to guest investigators\footnote{http://www.galex.caltech.edu/researcher/techdoc-ch2.html}.
\\
\\
The primary goal of \emph{GALEX} was study of the cosmic history of star formation as evidenced in the UV fluxes from star forming regions in distant galaxies.  In addition to this goal, many other much nearer UV sources were observed with great sensitivity over a large fraction of the sky.  Of particular interest here are \emph{GALEX} observations of white dwarfs (WDs).  .  WDs initially form as very hot (\textasciitilde $10^5$K) dense stellar cores, and as they cool, the bulk of their luminosity shifts from the extreme and far ultraviolet into the near ultraviolet and optical bands.  \emph{GALEX} can detect these stars at distances of several kiloparsecs.
\\
\\
One practical use of WDs, especially pure hydrogen DA WDs, is as flux calibration standards that span the wavelength range from the extreme ultraviolet into the near infrared (Sing et al. 2002).  Several properties of DA WDs favour their wide use as photometric and spectrometric flux standards in the UV and optical.  These are; 1) their spectral energy distributions are continuum-dominated, 2) the atmosphere of these stars are fully radiative and stable over a wide range of temperatures, 3) the opacities are due almost exclusively to neutral and ionized hydrogen and are thus very well determined, 4) the emergent stellar fluxes depend only on $T_{eff}$ and log\,g which can be determined spectroscopically, and 5) they are relatively nearby and free from interstellar reddening. See (Holberg \& Bergeron 2006) for a more detailed discussion of these points.
\\
\\
The \emph{GALEX} team originally used six WDs that had been designated by the \emph{Hubble Space Telescope} (\emph{HST}) as calibration standards.  However, all but one of these WDs caused saturation of both \emph{GALEX}'s detectors (Morrissey et al. 2007).  The DB WD LDS749b was used as a primary absolute flux calibrator for \emph{GALEX}.  However, following the Morrissey et al. (2007) results, Bohlin \& Koester (2008) published a refined model of this star on the \emph{HST} flux scale.  Specifically, these authors included a detailed spectroscopic analysis of the He I lines to set the $T_{eff}$ and log g ($T_{eff}$ =13,575 $\pm$ 50 K, and log g = 8.05 $\pm$ 0.7) and evaluated the effect of possible uncertainties in these parameters as well as interstellar reddening on the UV and IR fluxes for this star.  They concluded that this model now approaches the fidelity used to establish the three brighter DA white dwarfs (GD 71, GD 153, and G191B2B) as primary \emph{HST} flux standards.
\\
\\
The ability to precisely relate \emph{GALEX} fluxes to absolute fluxes in other bands is useful in several respects.  For example, \emph{GALEX}'s broad band UV fluxes can be used to identify stars having UV excesses, due to perhaps hot white dwarfs and/or subdwarfs (Bianchi et al., 2011).  The \emph{GALEX} bands are also more sensitive to low levels of interstellar extinction than other optical and near infrared broad band survey data, for example SDSS \emph{ugriz} colors.  Finally, for many objects it is useful to include \emph{GALEX} fluxes as contributions to multi-band spectral energy distributions.  All of these objectives are enhanced if \emph{GALEX} fluxes are on the same absolute flux scale as other observations.

\subsection{\emph{GALEX} Instrumentation}
The \emph{GALEX} telescope employs a Ritchey-Chr\'etin optical design, with a 50 cm diameter primary mirror.  In the focal plane light passes through an imaging window, a dispersive grism, or an opaque shutter, controlled by an optical wheel mechanism.  Incoming light also passes through a dichroic beam splitter, and is directed toward the  Near Ultraviolet (NUV, 1771-2831 \AA) and Far Ultraviolet (FUV, 1344-1786 \AA) detectors.  The beam splitter coating is chromatically selective, with a mean transmission 83 per cent in the NUV band, and a mean reflection of 61 per cent in the FUV band.  The transmitted light is reflected from the red blocking flat mirror to reduce the noise from zodiacal light ($\lambda > 3000 \AA$), and the reflected light passes through a multilayer filter that removes short wavelength geocoronal Ly$\alpha$ emissions (1216 \AA), as well as terrestrial OI airglow (1301-1356 \AA) \footnote{http://www.galex.caltech.edu/researcher/techdoc-ch1.html}.
\\
\\
The detectors are a pair of large format microchannel plate detectors (MCP)\footnote{http://www.galex.caltech.edu/researcher/techdoc-ch1.html}.  Each consists of a stack of three microchannel plates separating a photocathode and a delay line detecting anode.  Both MCP stacks are operated at a gain on the order of $1\times 10^7 - 2\times 10^7$, with operating voltages of 5200 and 6200V for the NUV and FUV detectors respectively (Jelinsky 2003).  MCP detectors were selected for \emph{GALEX} for their low background noise, high red rejection, and lack of cooling requirement.  Unfortunately, MCP detectors do have a lower quantum efficiency (around 8 per cent, depending on wavelength), as well as poor field flatness as compared to the more standard CCD detectors.  To mitigate local flatness variations, \emph{GALEX} moves its optical axis in a 1' spiral dither pattern, at a rate of approximately 1/2 rotation per minute.  This spreads the image of a point source over multiple 1.5" pixels so that pixel-to-pixel variance is averaged.  With dithering, the magnitude uncertainty of measured objects decreased from 0.068 and 0.125 mag in NUV and FUV to 0.027 and 0.050 mag in NUV and FUV, Morrissey et al. (2007).

\subsection{\emph{GALEX} Data Reduction and Calibration}
For point sources the procedures used to reduce \emph{GALEX} detector counts to extracted and calibrated point source images in celestial coordinates are described in detail in Morrissey et al. (2007). The  \emph{GALEX} data releases are referenced to the \emph{HST} photometric system of Bohlin et al. (2001) and Bohlin et al. (2008) through the defined spectrophotometric fluxes for a limited set of \emph{HST} reference standard WD stars.   Because \emph{GALEX} is a relatively sensitive broad band photometric instrument, many of these \emph{HST} standards are effectively too bright for \emph{GALEX}, particularly in the FUV detector.   The primary photometric standard for direct imaging used by \emph{GALEX} is the \emph{HST} standard LDS749b. LDS749b (WD2129+000) is a V= 14.674 DB (pure helium) WD which has band pass defined \emph{GALEX} AB magnitudes of 14.71 and 15.57 in the NUV and FUV, respectively (Morrissey et al. 2007).   The current and final \emph{GALEX} data release (GR7) still basically relies on this photometric calibration.  The more recent spectral model from Bohlin \& Koester (2008) gives magnitudes of 14.76 and 15.6 for NUV and FUV respectively.  The published \emph{GALEX} magnitudes for this star are 14.82 and 15.67 in NUV and FUV respectively.

\subsection{Objective}
The primary objective of this paper is to describe a simple transformation that can be applied to observed \emph{GALEX} GR7 archive NUV and FUV magnitudes that place them on the \emph{HST} absolute flux scale.  Although we discuss \emph{GALEX} detector effects such as saturation and dead time corrections, we rely mainly on empirical  correlations between observed \emph{GALEX} magnitudes and synthetic magnitudes to define 'corrected' fluxes.  Simple corrections of the type described here should be useful in directly transforming observed magnitudes to the \emph{HST} absolute scale, without the need for an additional layer of analysis of detector behavior.  In section 2 we discuss our methods of synthetic photometry.  In section 3, comparisons from synthetic photometry methods are described, and in section 4 we describe our corrections to \emph{GALEX} photometry.  Our conclusions are given in section 5.

\section{Procedure}

\subsection{Post GR7 \emph{GALEX} Calibration Based on White Dwarf Synthetic Magnitudes}
In this paper we follow the procedures of using synthetic photometry of DA white dwarfs to evaluate various widely used photometric systems and to place them on the \emph{HST} photometric system (Holberg \& Bergeron 2006).   In Holberg \& Bergeron synthetic fluxes were computed for large samples of DA white dwarfs which have spectroscopically determined effective temperatures and surface gravities based on the detailed fitting of observed H I Balmer lines.  The resulting models' energy distributions are then normalized to an observed flux, for example, a Johnson V or SDSS \emph{g} magnitude with respect to synthetic magnitudes computed for other bands. Holberg \& Bergeron also detail how this sort of synthetic DA photometry can be directly linked to the \emph{HST} photometric scale.   This makes possible a detailed comparison of observed magnitudes as a function of synthetic magnitudes and a natural way to define various photometric systems with respect to the \emph{HST} system in terms of a consistent set of photometric zero point fluxes.  In the present paper we apply these techniques to the \emph{GALEX} GR7 data set as it is currently defined.  Specifically, we use a significant sample of DA white dwarfs that have well-defined $T_{eff}$ and log g values, or alternately have well-determined \emph{International Ultraviolet Explorer} (\emph{IUE}) fluxes, to compute synthetic GALEX magnitudes and compare these to the corresponding observed \emph{GALEX} fluxes.   Using these techniques we can cover a wide range of \emph{GALEX} magnitudes, from the brightest unsaturated WDs to stars having AB magnitudes of \textasciitilde17. We investigate the residual linearity of \emph{GALEX} GR7 fluxes as well as small systematic offsets to the basic \emph{GALEX} calibration.

\subsection{Determination of White Dwarf Synthetic Magnitudes}
Two complimentary methods were used to compare observed \emph{GALEX} magnitudes with synthesized NUV and FUV magnitudes (Holberg \& Bergeron 2006).  The first employed model atmospheres computed from spectroscopically determined temperatures and gravities.  The second directly used the \emph{IUE} measured continua of WDs.  In each method, synthesized \emph{GALEX} magnitudes are calculated by the following:
\\
\begin{equation}
f_S=\frac{\int S(\lambda)f(\lambda)\lambda dlog(\lambda)}{\int S(\lambda)\lambda dlog(\lambda)}
\end{equation}

Where $f(\lambda)$ is the normalized model flux of the WD in $ergs\,cm^{-2} s^{-1} \AA^{-1}$, $S(\lambda)$ is the effective area of the instrument in $cm^2$, and $f_S$ is the integrated energy flux in $ergs\,cm^{-2} s^{-2}$.  The integrated flux can be converted into AB magnitudes by, 
\\
\begin{equation}
M_{AB} = -2.5 log(f_S) - 48.60
\end{equation}

The \emph{GALEX} photometric data of the WDs was gathered from the Mikulski Archive for Space Telescopes (MAST) public access site, using \emph{GALEX} GR7 pipeline data \footnote{http://galex.stsci.edu/GR6/}.  The coordinates of each of the target WDs were obtained from the Villanova Catalogue of Spectroscopically Identified White Dwarfs, McCook \& Sion (1999) \footnote{http://www.astronomy.villanova.edu/WDCatalog/}  and used to locate the objects.  For each WD, information on exposure time, published magnitude, and Poisson magnitude uncertainty were obtained for both the NUV and FUV detectors.  Where the Poisson uncertainty was less than the flat field uncertainty (0.027$mag_{AB}$ in NUV and 0.050$mag_{AB}$ in FUV, Morrissey et al. 2007), the flat field uncertainty was used instead.
\\
\\
Synthetic magnitudes were compared to the measured magnitudes.  Linear, quadratic, and cubic chi-squared minimizing interpolations were calculated for each set of data.  In all data sets, the quadratic interpolation had significantly more predictive power than the linear interpolation, while having only marginally less predictive power than the cubic interpolation.  Thus, the quadratic interpolations were ultimately used to calibrate the detectors.  

\subsection{Calibration With Respect to Model White Dwarf Atmospheres}
The first step in validating the \emph{GALEX} calibration using atmospheric models was to select a group of WDs whose spectra are easy to model.  The WDs selected were among those from Holberg \& Bergeron (2006) and Gianninas, Bergeron \& Ruiz (2011)that had been observed by \emph{GALEX}.  The WD sample was further reduced by applying the following criteria: 1) a high enough temperature that they emit significant flux over the entire \emph{GALEX} range (at least 13,000 K), 2) observed by the Sloan Digital Sky Survey (SDSS) and SDSS photometric data consistent with model spectra of an isolated white dwarf, 3) faint enough that they will not overwhelmingly saturate the \emph{GALEX} MCP detectors (no brighter than 11 mag), and 4) near enough that interstellar reddening is not a concern.
\\
\\
Using the spectroscopic $T_{eff}$ and log g from Gianninas et al. (2011), an unnormalized model spectrum of each of the WDs was calculated.  These model spectra extended from 1350 to 17000 \AA, containing both the Far Ultraviolet (FUV) band of \emph{GALEX} and the SDSS \emph{z} infrared band.  To normalize this spectrum, DR9 SDSS \emph{ugriz} photometry was used\footnote{http://skyserver.sdss3.org/dr9/en/tools/crossid/crossid.asp}.  Using the technique from Holberg \& Bergeron (2006), the photometric magnitudes from \emph{GALEX} and SDSS were both normalized to the \emph{HST} scale, and their uncertianties were calculated.  A normalization factor was chosen to minimize $\chi^2$ error from these photometric magnitudes.  When the initial model magnitudes were first compared to the measured magnitudes published by \emph{GALEX}, a general trend could be seen, but there were numerous significant outliers.  
\\
\\
One source of outliers was due stars that were too cool to have significant flux in the ultraviolet bands of \emph{GALEX}, and thus were dominated by background.  This lead to a practical temperature minimum cutoff of 13,000 K (i.e. spectral types of DA3.8 or earlier).  
\\
\\
A second source of outliers came from stars that were members of known binary systems.  The companions of these WDs although cooler, were luminous enough so that they had red and near IR excesses.  This excess light made normalization with respect to SDSS \emph{ugriz} photometry impossible.  For example, the star WD0232+035 (Feige 24) is a known binary system consisting of a white dwarf and an M-type star.  The plot in figure 1 shows a calculated spectrum from its atmospheric conditions.  The two points toward the left of the plot in figure 1 show the flux measured by the \emph{GALEX} FUV and NUV detectors; the remaining points are \emph{ugriz} photometric fluxes measured by the SDSS used to normalize the model spectrum.  The presence of the dMe companion results in an unmodeled red excess
\\
\\
\begin{figure}[h]
 \begin{center}
  \includegraphics[angle=90,width=80mm]{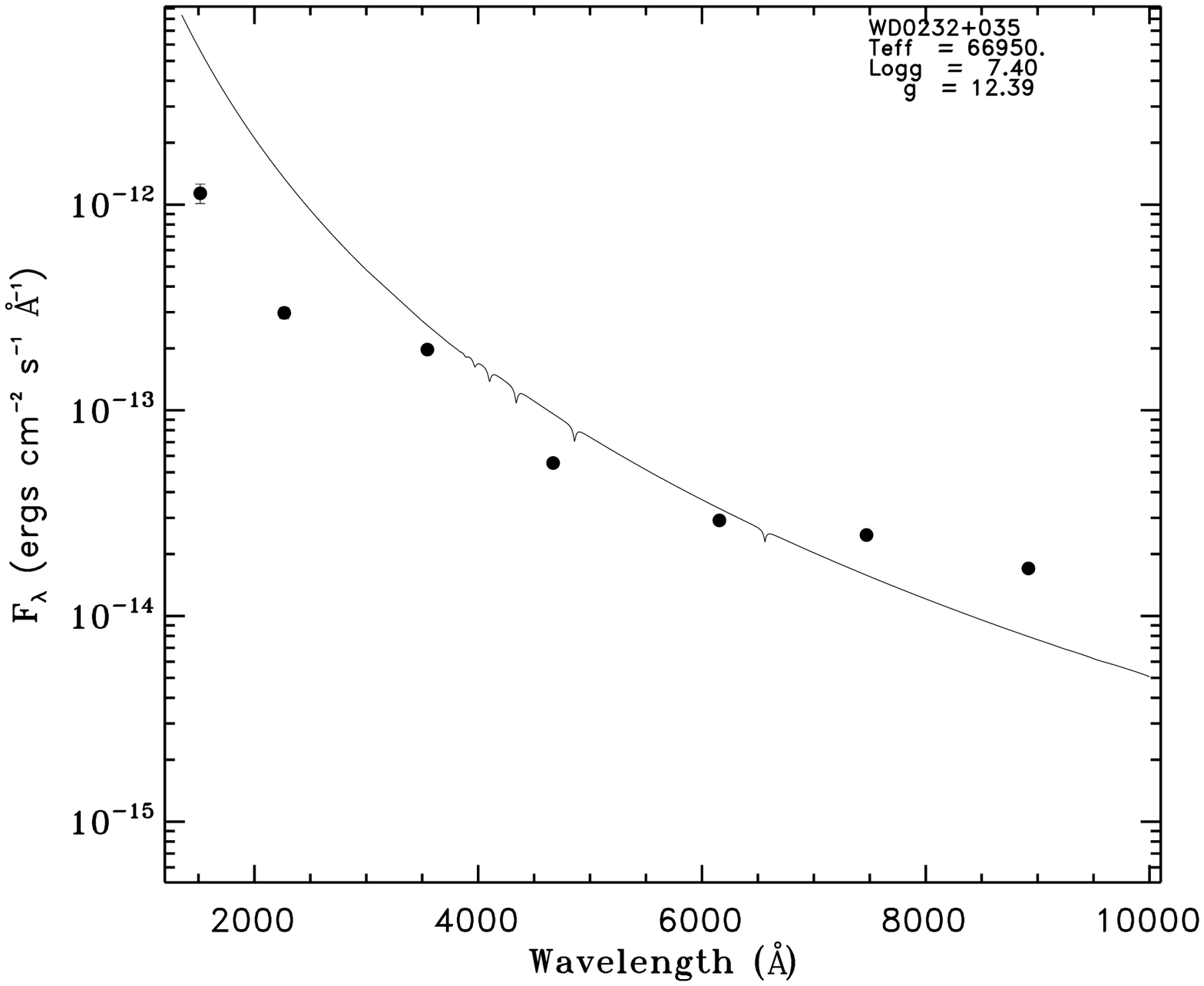}
  \includegraphics[angle=90,width=80mm]{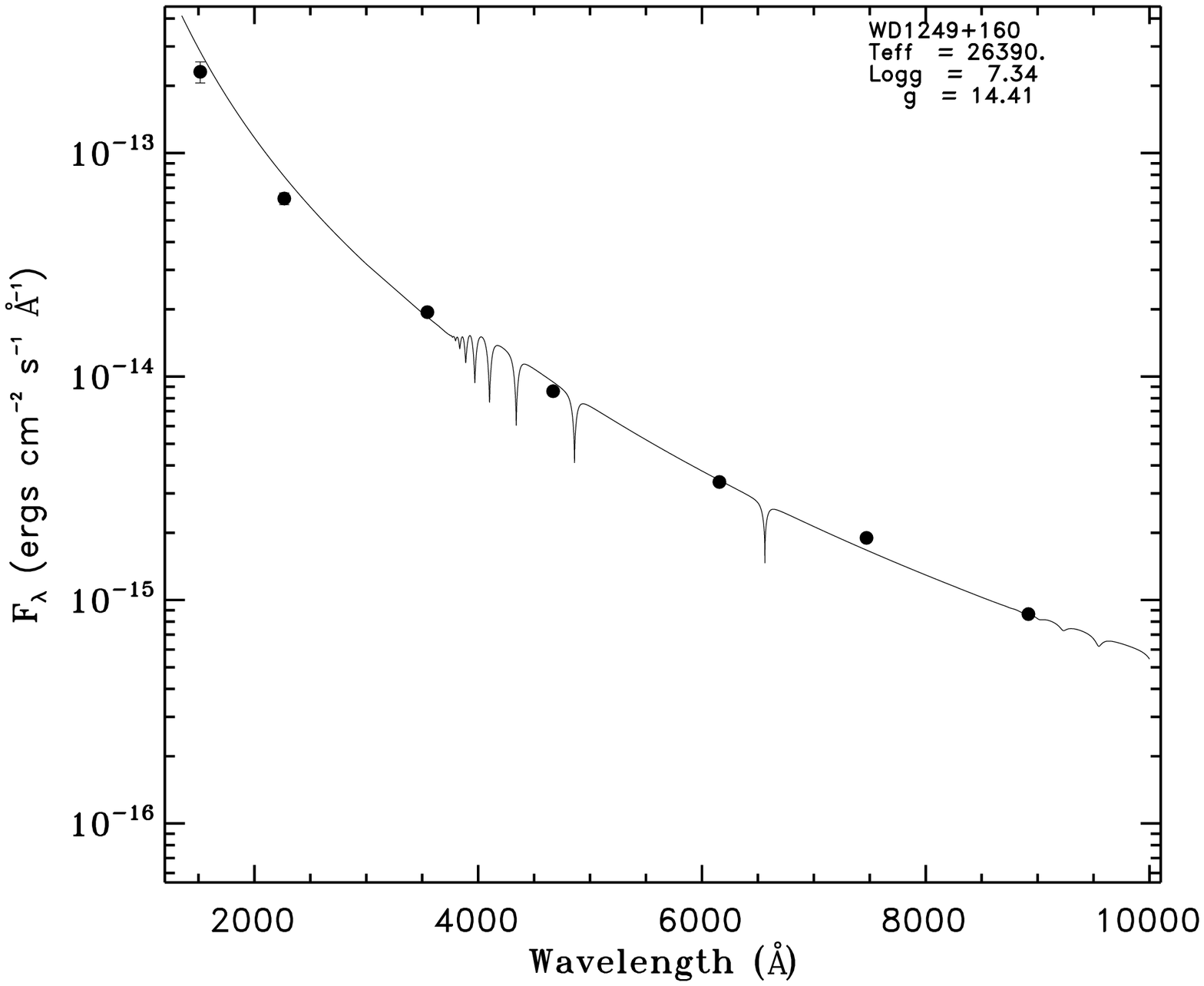}
  \caption{(Top) A model spectrum (curve) and SDSS \emph{ugriz} plus \emph{GALEX} photometry (points) for the well known WD+dMe binary system Feige 24 (WD0232+035).  The difficulty in normalizing SDSS photometry to the model due to thered excess of the secondary is evident in a binary system}
  \caption{(Bottom) A model spectrum (curve) and SDSS \emph{ugriz} plus \emph{GALEX} photometry (points) for the isolated WD (WD1249+160).  The model and photometric data are consistent for this star}
\end{center}
\end{figure}
\\
\\
For comparison, star WD1249+160 has no companion star.  The plot in figure 2 shows the same information as the plot in figure 1.  The \emph{ugriz} photometric fluxes match the model spectra flux.
\\
\\
A third source of outliers was from the detector itself. The initial measurement uncertainty was based solely on Poisson counting statistics.  However, the flat field uncertianties provided a much stronger source of uncertainty for most WDs.   Additionally, in 2003 the FUV detector began to exhibit an anomaly known as the 'blob' following a period of strong solar flares, which made some of the measurements questionable (Morrissey et al., 2007).  A few WDs were removed from consideration because they had been measured more than once with a large ($>3\sigma$) difference between the measurements.  WDs that were only measured once could not be tested this way, and may result in remaining outliers.  Additionally, communication with the \emph{GALEX} team has suggested that there are unmodeled sensitivity corrections that affects brighter stars near the edge of the detector field.  The data set that we used was insufficient to determine this correction, and it can be seen in a larger magnitude spread in the lower magnitude stars, on the order of ~0.1 magnitude uncertianty at 15th magnitude.
\\
\\
In figure 3 we show the empirical correlations between observed \emph{GALEX} NUV and FUV magnitudes and synthetic model fluxes.  The correlations do not match the anticipated one-to-one correlations (dashed lines) as there appears to be significant magnitude-dependent curvature for both data sets.  Quadratic and cubic expressions were fit to these data.  Over the range of magnitudes considered, quadratic fits gave reasonable representations of the data, while cubic fits did not significantly improve the fit.  We therefore have adopted the following expression to represent the best fit correlations between expected model magnitudes and observed \emph{GALEX} data:
\\
\begin{equation}
M_{GALEX} = c_0+c_1M_{syn} + c_2M_{syn}^2
\end{equation}

Where $M_{GALEX}$ and $M_{syn}$ are the observed\emph{GALEX} and synthetic AB magnitudes, and $c_0$, $c_1$, and $c_2$ are the respective quadratic coefficients.  The results for each band are given in table 1, along with the chi-squared per degree of freedom for each fit
\
\begin{figure}[h]
    \begin{center}
\includegraphics[angle=90,width=80mm]{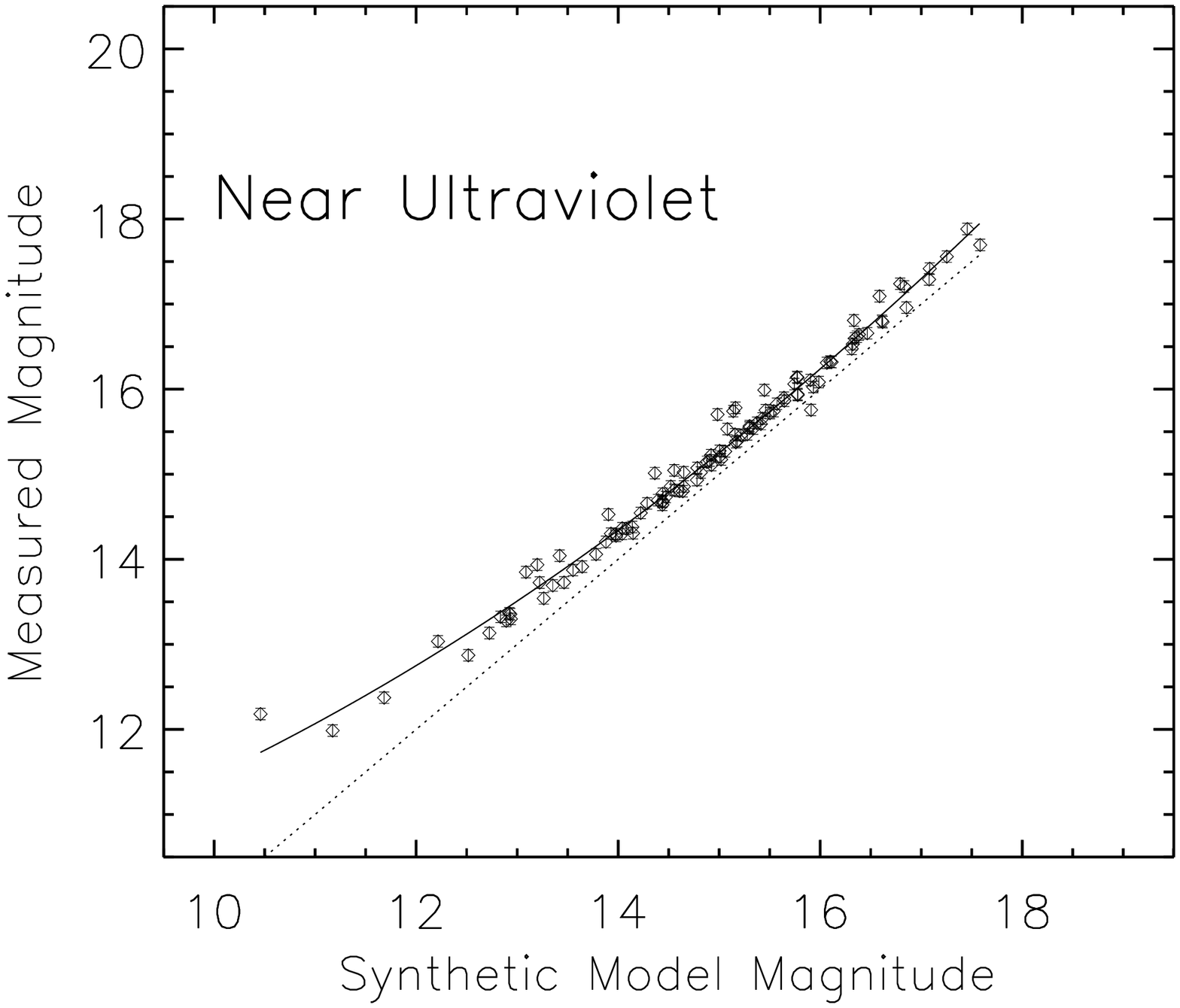}
\includegraphics[angle=90,width=80mm]{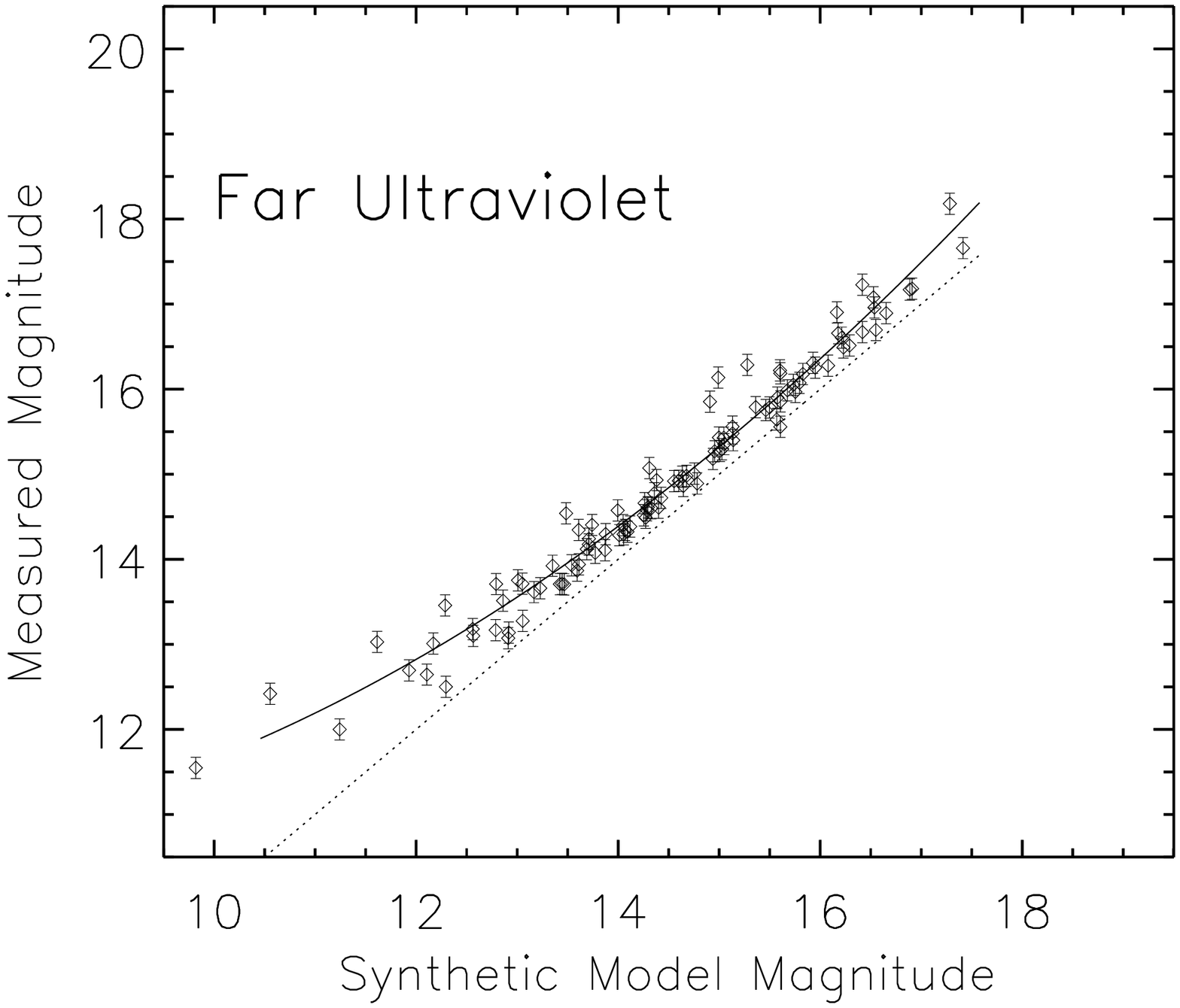}
\caption{The correlation between the observed \emph{GALEX} NUV (top) and FUV (bottom) magnitudes and synthetic magnitudes are shown.  The dashed lines are the expected one-to-one correlations, while the solid curves are quadratic fits to the data.}
    \end{center}
\end{figure}

\begin{center}
\begin{tabular} {c c c}
\multicolumn{3}{c}{Table 1}\\
\multicolumn{3}{c}{Quadratic fit parameters for atmospheric model}\\
\hline
\hline
Property & NUV & FUV\\
\hline
$c_0$ & 9.554 & 11.908\\
$c_1$ & -0.188 & -0.529\\
$c_2$ & 0.038 & 0.050\\
$\chi^2/dof$ & 4.81 & 3.531\\
number of stars & 107 & 99\\
Lower bound & 10.5 & 10.5\\
Upper bound & 17.5 & 17.5\\
\hline
\end{tabular}
\end{center}

The parameters of some of the stars used are shown below in table 2.  The remaining stars are available in the web version.\\
\small
\begin{center}
\begin{table*}
\begin{tabular} {c c c c c c c c c c c c}
\multicolumn{12}{c}{Table 2}\\
\multicolumn{12}{c}{Properties of WDs selected for model spectra calibration}\\
\hline
\hline
& &\multicolumn{2}{c}{Exposure Time}&\multicolumn{4}{c}{White Dwarf Properties}&\multicolumn{4}{c}{AB Magnitudes}\\
\cline{1-4}\cline{5-8}\cline{9-12}
\cline{1-4}\cline{5-8}\cline{9-12}\\
Star WDN& Survey & FUV & NUV & $T_{eff}$ & $T_{eff}$ err & Logg & Logg err & FUV & FUV err & NUV & NUV err\\
\hline
0000+171&AIS&172.2 & 178.2 & 21130 & 325 & 8 & 0.05 & 14.9807 & 0.125 & 15.2681 & 0.0675\\
0004+330&AIS & 112 & 465 & 49980 & 898 & 7.77 & 0.06 & 13.0285 & 0.125 & 13.0353 & 0.0675\\
0030-181&AIS & 126 & 126 & 14270 & 387 & 7.94 & 0.06 & 16.9028 & 0.125 & 16.808 & 0.0675\\
0058-044&AIS & 208 & 208 & 17370 & 292 & 8.1 & 0.05 & 14.8913 & 0.125 & 15.1735 & 0.0675\\
0127-050&AIS & 137 & 137 & 16790 & 250 & 7.99 & 0.04 & 14.6068 & 0.125 & 14.8028 & 0.0675\\
0129-205&AIS & 188.05 & 188.05 & 20670 & 333 & 8.01 & 0.05 & 14.4885 & 0.125 & 14.7995 & 0.0675\\
0155+069&AIS & 112 & 112 & 22840 & 355 & 7.84 & 0.05 & 14.2836 & 0.125 & 14.7707 & 0.0675\\
0231+050&AIS & 112 & 112 & 89470 & 5011 & 7.54 & 0.18 & 14.3451 & 0.125 & 14.6568 & 0.0675\\
\hline
\multicolumn{12}{c}{A machine readable version of the full table is available online}
\end{tabular}
\end{table*}
\end{center}
\normalsize

\subsection{Calibration With Respect to \emph{IUE} Spectra}
In a second calibration procedure, \emph{GALEX} WD measurements were compared to \emph{IUE} low dispersion spectrum.  The\emph{ IUE} SWP and LWP/LWR spectra camera observed 218 WD stars that were also observed by \emph{GALEX} during its operational lifetime.  These spectra were processed with NEWSIPS corrections and coadded where possible to produce observed UV spectral energy distributions between 1150 and 3000 \AA (Holberg, Barstow \& Burleigh 2003).  These previously measured spectra include interstellar extinction, and do not rely on model fluxes.  The downside of using the \emph{IUE} spectra is that they were not as well suited for this kind of analysis.  The observed spectra occasionally did not cover the entire detector pass band.  Finally, many of the measured spectra included had very large signal-to-noise ratios in the NUV and FUV ranges.  \emph{IUE} spectra used were obtained from the library of \emph{IUE} low dispersion spectra of white dwarfs \footnote{http://vega.lpl.arizona.edu/newsips/low/}, which were processed with \emph{IUE} NEWSIPS procedures and were effectively on the \emph{HST} flux scale (Holberg et al., 2003)
\\
\\
Both the best fit linear and best fit quadratic interpolations were calculated for both the FUV and NUV detectors.  For both detectors, the quadratic interpolations were significantly better than the linear interpolations, and are shown in figure 4.  The parameters for the the correlation are shown in Table 3
\\
\begin{figure}[h]
\begin{center}
\includegraphics[angle=90,width=80mm]{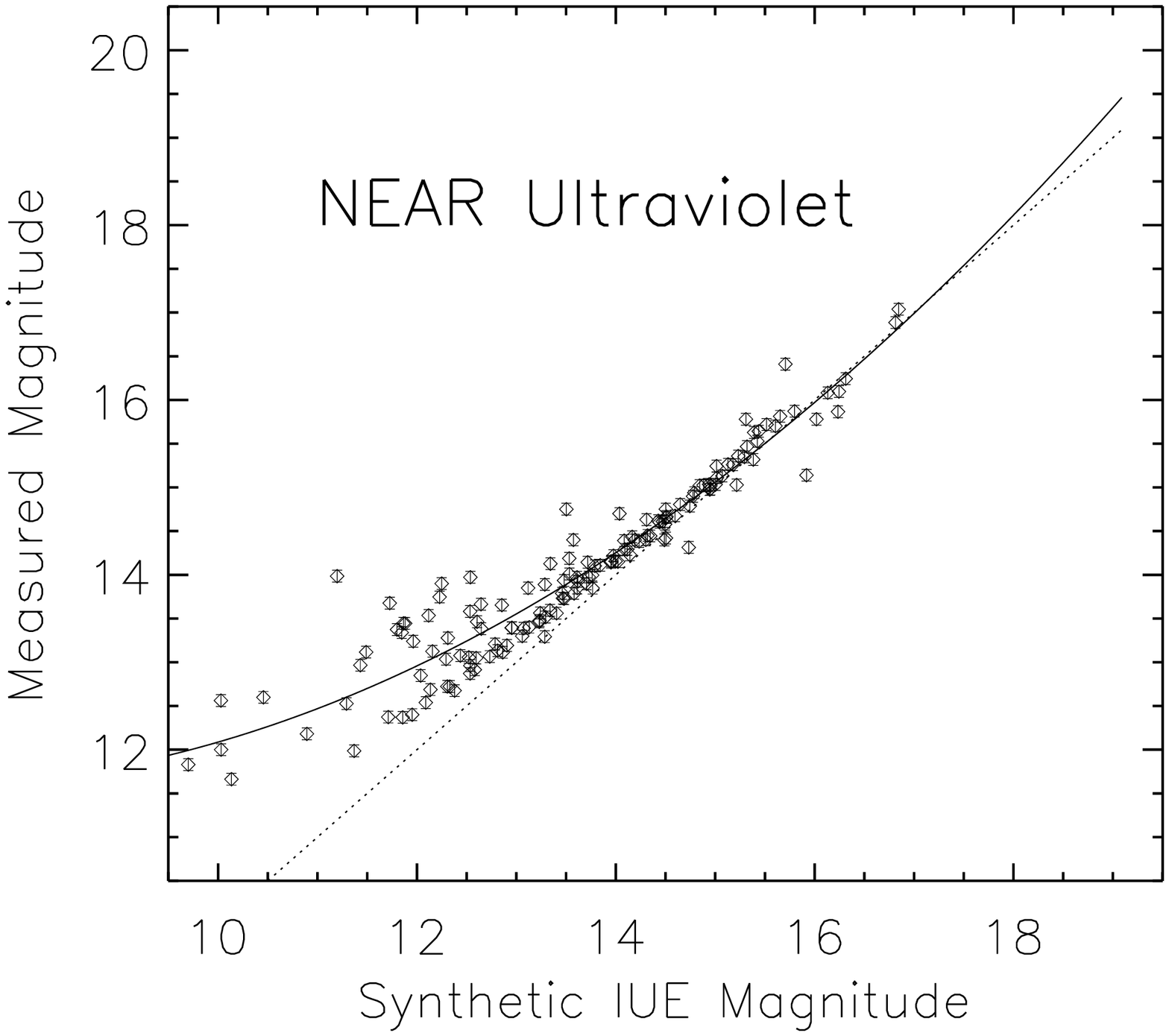}
\includegraphics[angle=90,width=80mm]{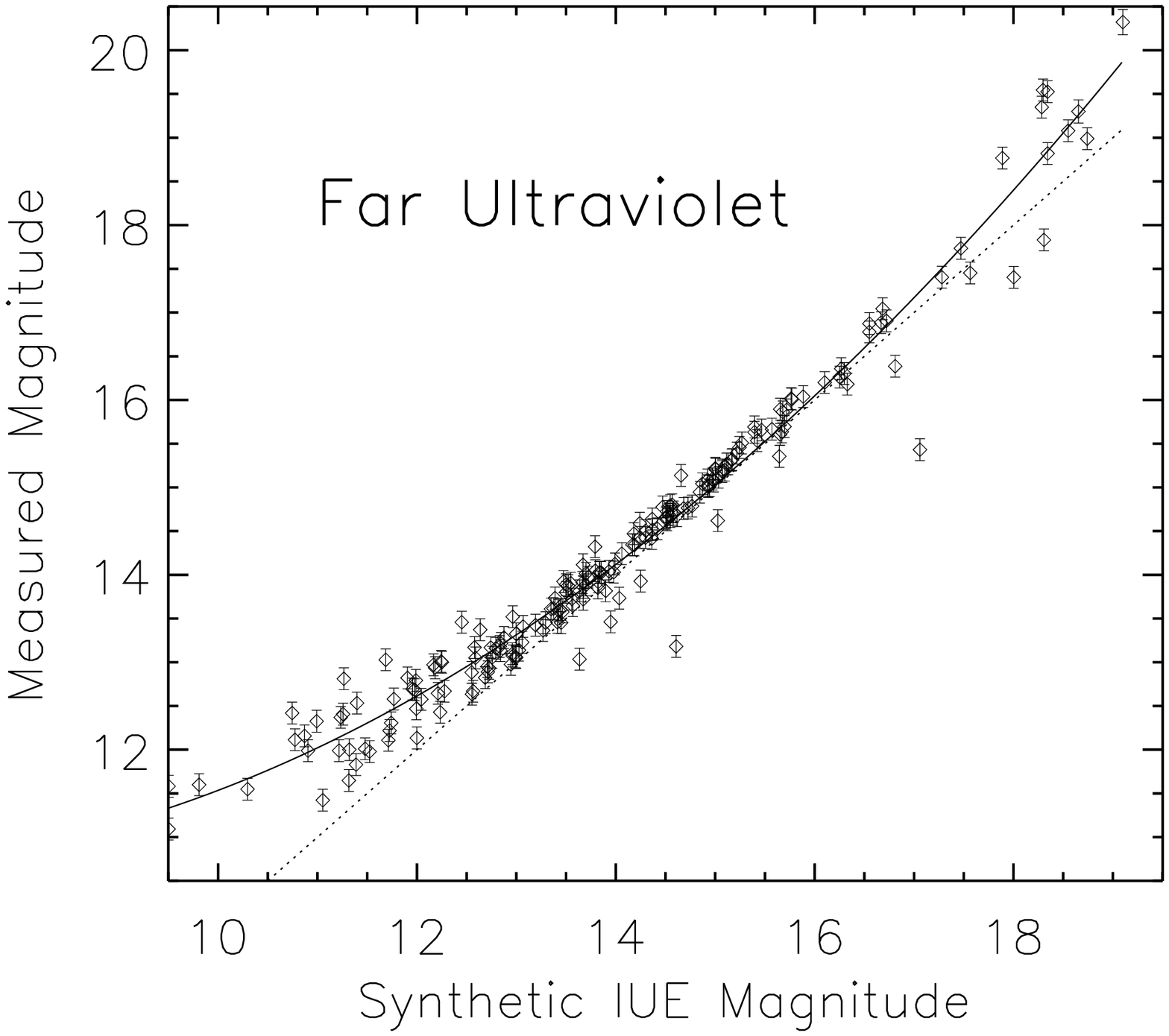}
\caption{The correlation between the observed \emph{GALEX} NUV (top) and FUV(bottom) magnitudes and predicted \emph{IUE} magnitudes are shown.  The dashed lines are the expected one-to-one correlations, while the solid curves are quadratic fits to the data}
\end{center}
\end{figure}
\\
\begin{center}
\begin{tabular} {c c c}
\multicolumn{3}{c}{Table 3}\\
\multicolumn{3}{c}{Quadratic fit parameters for IUE measurements}\\
\hline
\hline
Property & NUV & FUV\\
\hline
$c_0$ & 14.821 & 12.498\\
$c_1$ & -0.729 & -0.627\\
$c_2$ & 0.053 & 0.053\\
$\chi^2/dof$ & 16.807 & 5.989\\
number of stars & 197 & 218\\
Lower bound & 9.5 & 10.5\\
Upper bound & 17.5 & 17.5\\
\hline
\end{tabular}
\end{center}

The properties of some of the stars used can be seen below in table 4.  The remaining stars are available in the web version.\\
\small
\begin{center}
\begin{table*}
\begin{tabular} {c c c c c c c c}
\multicolumn{8}{c}{Table 4}\\
\multicolumn{8}{c}{Properties of WDs selected for IUE spectra calibration}\\
\hline
\hline
&&\multicolumn{2}{c}{Exposure Time}&\multicolumn{4}{c}{AB Magnitudes}\\
WDN & Survey & FUV & NUV & FUV & FUV err & NUV & NUV err\\
\hline
0000-170 & AIS & - & 63.05 & - & - & 14.8972 & 0.0675\\
0002+729 & AIS & 208 & 208 & 15.2669 & 0.125 & 14.6308 & 0.0675\\
0004+330 & AIS & 112 & 465 & 13.0285 & 0.125 & 13.0353 & 0.0675\\
0005+511 & AIS & 219 & 219 & 12.3251 & 0.125 & 13.2181 & 0.0675\\
0017+136 & AIS & - & 122 & - & - & 14.9812 & 0.0675\\
0022-745 & AIS & 202 & 202 & 14.0467 & 0.125 & 14.3996 & 0.0675\\
0037+312 & AIS & 109 & 109 & 13.2827 & 0.125 & 13.5633 & 0.0675\\
0041+092 & AIS & 179.05 & 179.05 & 13.1606 & 0.125 & 13.4619 & 0.0675\\
\hline
\multicolumn{8}{c}{A machine readable version of the full table is available online}
\end{tabular}
\end{table*}
\end{center}

\section{Comparison of Synthetic Magnitude Methods}
There were a total of 19 WDs that were used in both the model atmosphere method and the \emph{IUE} method.  All 19 WDs have synthetic FUV magnitudes from both methods, but only 11 have synthetic NUV magnitudes from available \emph{IUE} data.  The atmospheric model synthetic magnitudes averaged 0.168 $mag_{AB}$ higher with a standard deviation of 0.0627 $mag_{AB}$ in FUV, and 0.580 $mag_{AB}$ lower with a standard deviation of 0.356 $mag_{AB}$ in NUV.  Considering all 32 measurements, the atmospheric model synthetic magnitudes averaged 0.300 $mag_{AB}$ lower with a standard deviation of 0.376 $mag_{AB}$.  For the individual stars, the atmospheric model spectrum and the \emph{IUE} spectrum are remarkably similar.  For example, the spectrum of star WD1658+440 (PG1658+440) is shown in figure 5, along with the response functions of the detectors.  For this particular star, the total difference between the atmospheric spectrum and the \emph{IUE} spectrum is 2.8 per cent, with an RMS of 9.8 per cent.  Most of the difference in the spectrum occurs outside of the band that the detectors are sensitive to; eliminating that part, the total difference becomes just 1.5 per cent.
\\
\begin{figure}[h]
\begin{center}
\includegraphics[angle=90,width=90mm]{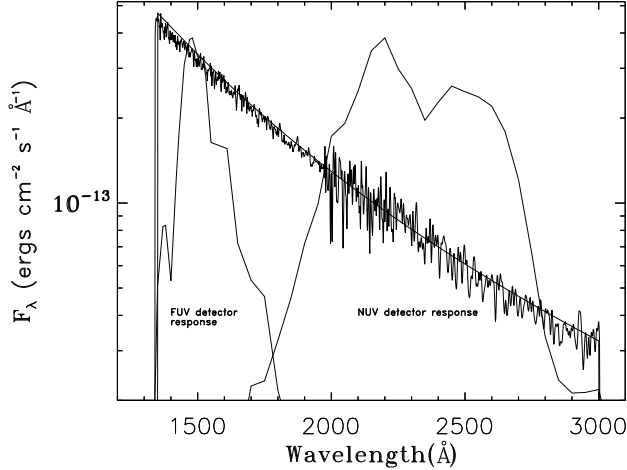}
\caption{Atmospheric model flux and \emph{IUE} flux.  Relative FUV and NUV detector response functions}
\end{center}
\end{figure}
\\
When looking at all 11, these results were typical.  Most atmospheric model and \emph{IUE} comparisons had total flux differences on the order of 5 per cent, with RMS values on the order of 7 per cent.  Two stars had average flux differences greater than 10per cent, WD0343-007 (KUV0343-007) and WD0939+262 (PG0939+262).  Both WDs were outliers in the \emph{IUE} synthetic flux model.

\section{Calibration}
\subsection{Corrections to \emph{GALEX} GR7 Magnitudes}
The correlations between observed \emph{GALEX} GR7 magnitudes and our model-based and \emph{IUE}-based synthetic magnitudes are similar, indicating that a similar effect is being measured.  We have selected the model-based synthetic magnitudes to define our empirical corrections to the \emph{GALEX} AB magnitudes.  This is due to the fact that our synthetic magnitudes have lower inherent dispersion with respect to the quadratic correlations (see figure 3) than the \emph{IUE} correlations (figure 4)
\\
\\
The expressions that convert observed \emph{GALEX} AB magnitudes into 'corrected' \emph{GALEX} magnitudes as a function of observed \emph{GALEX} magnitudes are basically an inversion of the correlations in figure 3.  We have chosen explicit quadratic solutions to equations to express these relations.  An alternate method that reverses the correlation plots in Fig. 3 and refits the data to observed \emph{GALEX} magnitudes is also possible, but represents the data with slightly less fidelity.  Our correlations to the \emph{GALEX} magnitudes are :
\begin{equation}
M_{corr} = C_0 + (C_1M_{obs} + C_2)^{1/2}
\end{equation}

The $M_{corr}$ and $M_{obs}$ are the respective corrected and observed \emph{GALEX} magnitudes, and $C_0$, $C_1$, and $C_2$ are calculated constants.  These values were calculated by inverting the quadratic best fit lines from the synthtic magnitudes and then slightly adjuting the constant $C_0$ s that the centroid of the frequency distribution of residuals was zero.  The corresponding values of these constants for the NUV and FUV corrections are given in table 5.  The magnitude ranges over which these expressions are valid are also contained in Table 5.
\begin{center}
\begin{tabular} {c c c}
\multicolumn{3}{c}{Table 5}\\
\multicolumn{3}{c}{Inverse Quadratic Corrections}\\
\hline
\hline
Property & NUV & FUV\\
\hline
$C_0$ & 2.634 & 5.371\\
$C_1$ & 26.316 & 20.000\\
$C_2$ & -245.329 & -210.200\\
number of stars used & 293 & 298\\
lower measured magnitude limit & 9.321 & 10.509\\
upper measured magnitude limit & 17.5 & 17.5\\
$\chi^2/dof$ & 3.824 & 4.623\\
\hline
\end{tabular}
\end{center}
The results of applying our recalibration corrections to the measured \emph{GALEX} magnitudes are seen in figure 6.  In the NUV, the average difference between the synthetic magnitude and the re-calibrated magnitude was 0.003$mag_{AB}$ with a variance of 0.154 $m_{AB}$.  In the FUV, the average difference between the synthetic magnitude and the calibrated magnitude was 0.002 $mag_{AB}$ with a variance of 0.134 $mag_{AB}$.  In both bands, the average difference did not vary with magnitude, but the variance of the difference was largest at lower magnitudes.
\\
\begin{figure}[h]
\begin{center}
\includegraphics[angle=90,width=80mm]{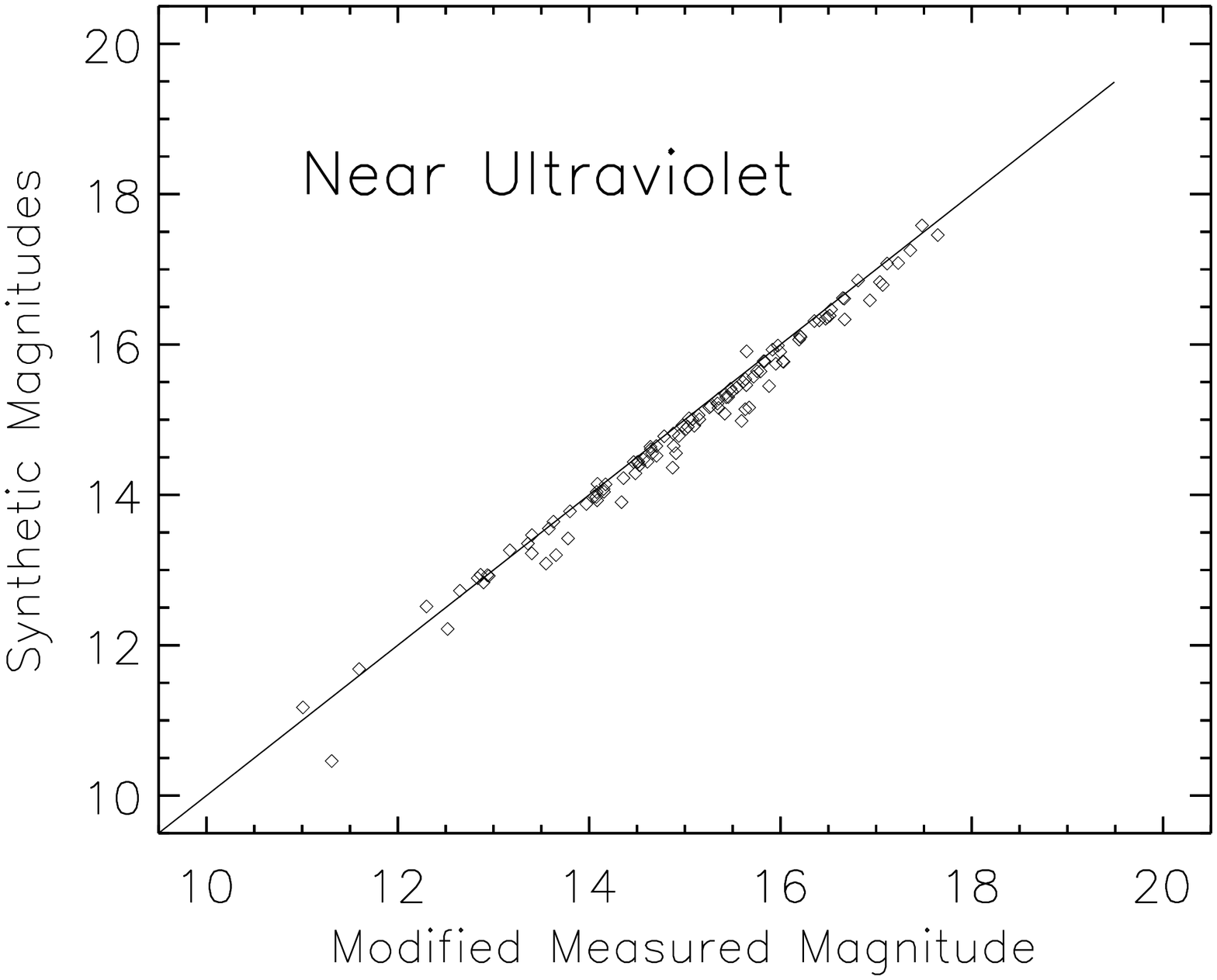}
\includegraphics[angle=90,width=80mm]{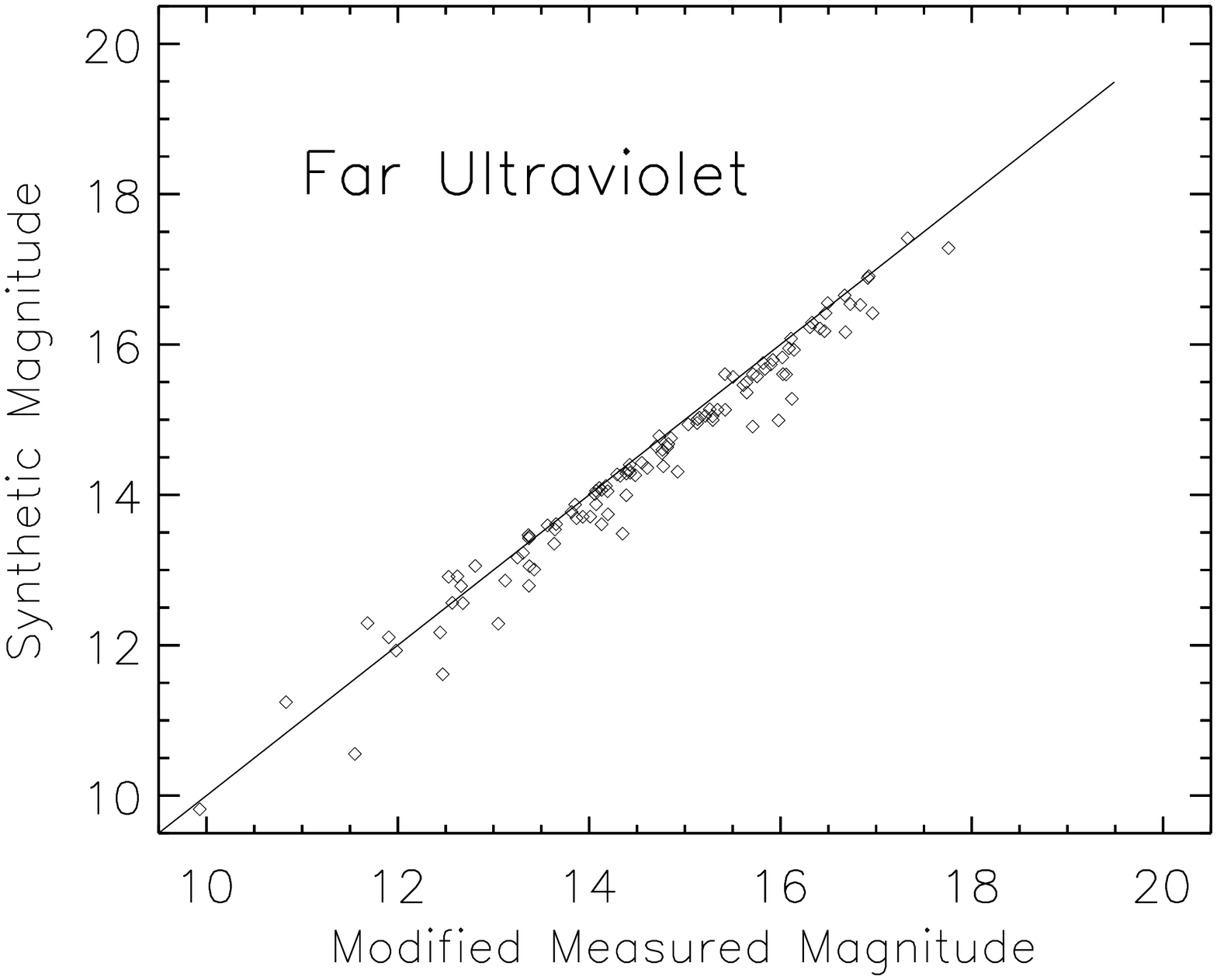}
\caption{Comparison of synthetic model magnitudes and corrected measured magnitudes for NUV (top) and FUV (bottom).  The lines show the expected one-to-one correlation.}
\end{center}
\end{figure}
\\
\\
\\
One surprising result of these calibration calculations is the correction at higher magnitudes.  Typically, non-linearities in measurement occur primarily in brighter stars with very high measured count rates.  Figures 3 and 4 do show a non-linearity in the fainter stars.  We hypothesize that this could be due to interstellar reddening, since fainter white dwarfs tend to be farther away.  However, it should be noted that there were only a few stars in this region, so corrections for stars dimmer than magnitude 16 should be treated with a little more caution.
\\
\\

\subsection{Dead Time Correction}
The most apparent non-linearity seen in figure 3 corresponds to the brightest stars.  This suggests effects associated with the local count rate behavior of the detector or perhaps with the aperture extractions.  Count rate related detector effects can be related to detector dead time.  The \emph{GALEX} images are corrected for global dead time so this effect should
be negligible.   On the other hand the ability of the detector to deal with high local (point source related) count rates are stated in Morrissey et al. (2007) to be less well understood and are difficult to investigate in detail.   Bright point sources can locally saturate the MCP reducing gain.  Morrissey et al. provide plots which measure of this effect for both detectors.  In our analysis we begin with the MAST GR7 magnitudes and empirically determine the correlation of observed magnitudes with predicted count rates.  In the absence of a detailed analysis of detector behavior this is perhaps the most effective way to estimate first order magnitude corrections.
\\
\\
Although the current limiting and gain reducing aspects of local high count rates are complicated, we have attempted to use our results in Fig. 3 to interpret the effects as a standard non-paralyzable 'dead time correction' , where the term non-paralyzable refers to time following the measurement an event the discriminator electronics are unable to register another event.   The expression for a dead time correction in terms of observed count rate $C_{obs}$ and true count rate $C_{true}$ for a dead time $\tau$ is given by:

\begin{equation}
C_{true} = \frac{C_{obs}}{1-C_{obs}*\tau}
\end{equation}

GALEX magnitudes can be estimated from count rates as:

\begin{equation}
M=-2.5log_{10}(C) + Z_p
\end{equation}

Where C is the count rate and $Z_p$ are the respective NUV and FUV zero point magnitudes 20.02 and 18.82 mag (Morrissey et al. 2007). If the correlations in Fig. 3 are fit for observed \emph{GALEX} magnitudes vs synthetic magnitudes and $\tau$ is found to minimize $\chi^2$ error, we find respective dead times of 578 and 1209 $\mu$s for the NUV and FUV data.  When these dead time corrections were applied to the data, the results were somewhat problematic.  Corrected magnitudes near the middle of the brightness range (12-14 mag) tended to be too low, while those for the brighter stars (9-11 mag) tended to be too high.  This suggests that the dead time had a larger effect on brighter stars than purely nonparalyzable dead time would, which suggests that the dead time is paralyzable or semiparalyzable. Local dead time most likely comes from the decrease in MCP voltage following a discharge, also known as gain sag.  In summary, a simple dead time correction model appears less than adequate.

\section{Conclusions}
We have conducted an empirical evaluation of the absolute calibration of \emph{GALEX} fluxes with respect to DA white dwarfs.  Two independent methods were used, the first involving model atmospheres normalized to SDSS \emph{ugriz} magnitudes and the second, a direct use of \emph{IUE} low dispersion spectra which were integrated with the \emph{GALEX} FUV and NUV camera response curves to synthesize predicted \emph{GALEX} magnitudes.  We determined that \emph{GALEX} fluxes possess modest magnitude-dependent departures from the expected one-to-one correlations with the predicted magnitudes.  We provide empirical quadratic magnitude-dependent corrections for observed GALEX magnitudes that place GALEX fluxes in better agreement with the AB magnitude scale.  The corrections described here, strictly speaking, pertain to point sources with flat or rising spectral energy distributions.   No investigations were made as to the applicability of these corrections to GALEX grism spectra or diffuse sources.
\\
\\
The GALEX corrections presented here have a faintness limit of approximately 17.5 $mag_{AB}$.  This is primarily due to the general lack of suitable fainter DA stars having robust measurements of $T_{eff}$ and log g. However, it is feasible to extend our corrections to magnitudes of 19 or 20 using SDSS DA white dwarfs.  Kleinman et al (2013) have presented a spectroscopic analysis of over 20,000 WD in the SDSS DR7 catalog.  This includes approximately 12000 DA WDs.   Use of these stars, however, will require a determination of interstellar reddening in both the SDSS bands and the GALEX bands.  Holberg, Bergeron \& Gianninas (2008) has shown that this is feasible to determine both distances and an interstellar reddening parameter (E(B-V) from the SDSS data for DA WDs.   Such a project would also require determining if there are a sufficient small number of representative Galactic reddening curves that can predict interstellar extinction in the GALEX bands.  Fortunately, the areal distribution of the Klienman et al. catalog is sufficiently dense (\textasciitilde several stars per sq. deg) that such a project does appear feasible.
\\
\\

\section{Acknowledgements}
Lawrence Camarota acknowledges support from the University of Arizona Department of Physics.  J.B.H. acknowledges support from NASA Astrophysics Data Program grant NNX1OAD76. We also wish to thank Patrick Morrissey for his many suggested improvements to an earlier draft of this paper.
\\
This research has made use of the WD Catalog maintained at Villanova University and the Some/all of the data presented in this paper were obtained from the Mikulski Archive for Space Telescopes (MAST). STScI is operated by the Association of Universities for Research in Astronomy, Inc., under NASA contract NAS5-26555. Support for MAST for non-HST data is provided by the NASA Office of Space Science via grant NNX09AF08G and by other grants and contracts
\\
Funding for the SDSS and SDSS-II has been provided by the Alfred P. Sloan Foundation, the Participating Institutions, the National Science Foundation, the U.S. Department of Energy, the National Aeronautics and Space Administration, the Japanese Monbukagakusho, the Max Planck Society, and the Higher Education Funding Council for England. The SDSS Web Site is http://www.sdss.org/.
\\
The SDSS is managed by the Astrophysical Research Consortium for the Participating Institutions. The Participating Institutions are the American Museum of Natural History, Astrophysical Institute Potsdam, University of Basel, University of Cambridge, Case Western Reserve University, University of Chicago, Drexel University, Fermilab, the Institute for Advanced Study, the Japan Participation Group, Johns Hopkins University, the Joint Institute for Nuclear Astrophysics, the Kavli Institute for Particle Astrophysics and Cosmology, the Korean Scientist Group, the Chinese Academy of Sciences (LAMOST), Los Alamos National Laboratory, the Max-Planck-Institute for Astronomy (MPIA), the Max-Planck-Institute for Astrophysics (MPA), New Mexico State University, Ohio State University, University of Pittsburgh, University of Portsmouth, Princeton University, the United States Naval Observatory, and the University of Washington.
\section{Reference}
Bianchi L. et al. 2000, Mon. Soc. Ast. Italiana, 71, 1117
\\
Bianchi L. et al. 2011 MNRAS, 411, 2770
\\
Bohlin R. C. 2001, AJ, 122, 2118
\\
Bohlin R. C. \& Koester D. 2008, ApJ, 135, 1098
\\
Gianninas A., Bergeron P., \& Ruiz M.T. 2011, ApJ, 743, 138
\\
Holberg J.B., Barstow M.A., \& Burleigh M.R. 2003, ApJS, 147, 145
\\
Holberg J.B.\& Bergeron P. 2006, AJ, 132, 1221
\\
Holberg J.B., Bergeron P. \& Gianninas A. 2008, AJ, 135,1225
\\
Jelinsky P. 2003, SPIE, 4854, 233
\\
Kleinman S.J. et al. 2013, ApJS, 204, 5
\\
McCook G. \& Sion E.M. 1999, ApJS, 121, 1
\\
Morrissey P. et al. 2007, ApJS, 173, 682
\\
Sing D., Holberg J.B., \& Dupuis J. 2002, ASP Conference Series, 264, 575

\end{document}